\documentclass[]{article}
\usepackage{textcomp}
\usepackage{wrapfig}
\usepackage{lineno,hyperref}
\usepackage{graphicx}
\usepackage{epstopdf}
\usepackage{tikz-cd}
\usepackage{array}
\usepackage{multirow}
\usepackage{amssymb}
\usepackage{amsmath}
\usepackage{textcomp}
\usepackage[euler]{textgreek}
\usepackage{nicefrac}
\usepackage{array}
\usepackage{booktabs}
\usepackage[margin=1.25in]{geometry}
\usepackage{xcolor}
\usepackage{authblk}
\usepackage{cancel}
\usepackage{tabularx}
\usepackage{makecell}
\usepackage{caption}
\usepackage{subcaption}

\usepackage{braket}
\usepackage[leftcaption]{sidecap}
\graphicspath{ {images/} }

\newsavebox{\measurebox}

\modulolinenumbers[5]

\date{}

\begin{document}

\title{Design and Performance of the ALPS\,II Regeneration Cavity}


\author[1]{Todd~Kozlowski\thanks{todd.kozlowski@desy.de}}
\author[1]{Li-Wei~Wei}
\author[1]{Aaron~D.~Spector}
\author[2]{Ayman~Hallal}
\author[1]{Henry~Fr\"adrich}
\author[3]{Daniel~C.~Brotherton}
\author[1]{Isabella~Oceano}
\author[2]{Aldo~Ejlli}
\author[4]{Hartmut~Grote}
\author[3]{Harold~Hollis}
\author[2]{Kanioar~Karan}
\author[2,3]{Guido~Mueller}
\author[3]{D.B.~Tanner}
\author[2]{Benno~Willke}
\author[1]{Axel~Lindner}
\affil[1]{Deutsches Elektronen-Synchrotron DESY, 22607 Hamburg, Germany}
\affil[2]{Max-Planck-Institut f\"ur Gravitationsphysik (Albert-Einstein-Institut) and Leibniz Universit\"at Hannover, 30167 Hannover, Germany}
\affil[3]{Department of Physics, University of Florida, 32611 Gainesville, Florida, USA}
\affil[4]{School of Physics and Astronomy, Cardiff University,  CF24 3AA Cardiff, Wales, United Kingdom}

\maketitle

\begin{abstract} 
The Regeneration Cavity (RC) is a critical component of the Any Light Particle Search\,II (ALPS\,II) experiment. It increases the signal from possible axions and axion-like particles in the experiment by nearly four orders of magnitude. The total round-trip optical losses of the power circulating in the cavity must be minimized in order to maximize the resonant enhancement of the cavity, which is an important figure of merit for ALPS\,II. Lower optical losses also increase the cavity storage time and with the 123 meter long ALPS II RC we have demonstrated the longest storage time of a two-mirror optical cavity. We  measured a storage time of $7.17\pm0.01\rm\,ms$, equivalent to a linewidth of $44.4\,\rm Hz$ and a finesse of 27,500 at a wavelength of 1064\,nm.

\end{abstract}

\newcommand{\murm}{%
  \ifmmode
    \mathchoice
        {\hbox{\normalsize\textmu}}
        {\hbox{\normalsize\textmu}}
        {\hbox{\scriptsize\textmu}}
        {\hbox{\tiny\textmu}}%
  \else
    \textmu
  \fi
}

\section{Introduction}
\begin{figure}
    \centering
    \includegraphics[width=\linewidth]{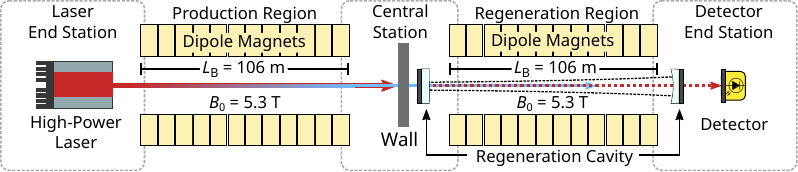}
    \caption{ALPS\;II conceptual layout for the first science campaign. A potential field of light (pseudo)scalars (blue) is generated from the high-power laser in the magnetic field of the production region. The field passes unimpeded through an optically opaque wall and into the Regeneration Cavity, where a small fraction of it is reconverted to light (dotted red) for detection.}
    \label{fig:alps2}
\end{figure}

Light-shining-through-a-wall (LSW) experiments aim to produce, and then subsequently detect, light bosons via their mixing with photons, where (pseudo-)scalar and tensor particles require a background magnetic field \cite{Sikivie1983,Anselm1985,vanbibber1987}. ALPS\,II \cite{diaz2022} is a LSW experiment located at the Deutsches Elektronen-Synchrotron DESY in Hamburg, Germany, which uses optical cavities to resonantly enhance the experimental sensitivity to such particles, including the axion \cite{Hoogeveen1990,Sikivie2007,Mueller2009}. In the design of the ALPS\,II first science campaign, shown in Figure \ref{fig:alps2}, light from a high-power continuous-wave laser in the laser end station enters the production region. The flux of axions generated by the interaction of the light with the magnetic field of the production region then enters the regeneration region of the experiment through an optically opaque `wall'. The purpose of the Regeneration Cavity (RC) is to subsequently enhance the reconversion rate of axions to light for detection. 

The wall and flat RC mirror are housed in the central station, with the RC curved mirror located at the detector end station on the opposite side of the regeneration region magnet string. Each magnet string consists of twelve superconducting HERA dipole magnets with field strength of 5.3\,T. In order to maximize the ALPS II sensitivity, the 123 meter long RC was constructed and aligned in such a way that the spatial mode of the axion field, which follows the spatial mode of the high-power laser as if it were unimpeded by the wall, is well matched to the fundamental mode of the optical cavity. For axions masses below 0.1\,meV, the regenerated photon rate of ALPS\,II for a given axion-photon coupling $g_{a \gamma \gamma}$, can be expressed as \cite{arias2010,redondo2011}
\begin{equation}\label{eq:signalrate}
    \dot{N}_{\mathrm{reg}} \approx \frac{1 \rm \, photon}{14 \rm \, hours} \cdot \left( \frac{P_{\mathrm{laser}}}{40 \, \rm W} \right) \cdot \left( \frac{\beta}{10,000} \right) \cdot \left( \frac{\eta}{0.9} \right) \cdot \left( \frac{g_{a \gamma \gamma}}{2 \times 10^{-10}\, \rm GeV^{-1}} \right)^4 \cdot \left( \frac{B_0}{5.3 \, \rm T} \right)^{4} \cdot \left( \frac{L_\text{B}}{106 \, \rm m} \right)^{4}
\end{equation}
where $\dot{N}_{\mathrm{reg}}$ is the regenerated signal rate in terms of photons per second, $P_\text{laser}$ is the power of the laser in the production area
, $B_0$ is the external magnetic field strength, and $L_\text{B}$ is the magnetic field length. The factor $\eta$ represents the coupling efficiency between the regenerated light field and the fundamental eigenmode of the RC in terms of power. $\beta$ is the resonant enhancement factor of the RC and is the principal figure of merit for the cavity's performance.\cite{diaz2022}.

For a high-finesse RC, the resonant enhancement factor can be expressed as
\begin{align}\label{eq:powerbuildup}
    \beta \approx \frac{4T_1}{(T_1 + T_2 + l)^2} 
\end{align}
where $T_1$ represents the transmissivity of the mirror through which regenerated light couples out of the cavity for detection, $T_2$ is the transmissivity of the other mirror, and $l$ is the excess optical loss, the fraction of power unintentionally lost from the cavity fundamental eigenmode in each round-trip. The minimization of this round-trip loss leads to improvement in the ALPS\,II photon regeneration rate. In terms of the cavity free spectral range $f_0$ and storage time $\tau_\text{s}$, $\beta$ can be expressed as
\begin{align}
  \beta \approx T_1 \cdot (f_0 \cdot \tau_\text{s})^2
\end{align}
In this paper we report on the ALPS\,II Regeneration Cavity which has achieved a storage time of $7.17\pm0.01\rm\,ms$, to the best of our knowledge the longest ever reported for a two-mirror Fabry-Perot optical cavity. This is equivalent to a $44.4\,\text{Hz}$ linewidth, a finesse of 27,500, and a resonant enhancement factor of 7600. We also discuss the long-term performance of the RC, necessary for the successful operation of ALPS\,II. In particular, we characterize the cavity in the context of conditions present for the ALPS II first science campaign, which took place from January 26th until May 5th, 2024. 

\section{Experimental Setup}\label{Sec:setup}

\begin{figure}
    \centering
    \includegraphics[width=\linewidth]{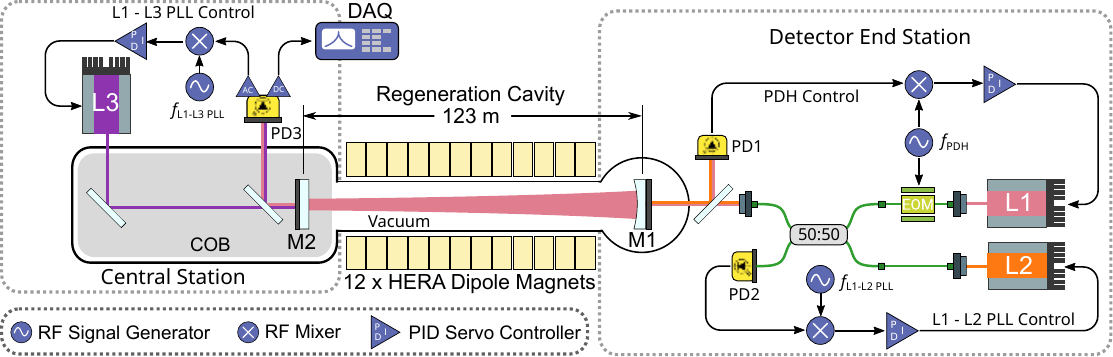}
    \caption{Schematic of the Regeneration Cavity and measurement setup. Not shown are optics responsible for the alignment, mode-matching, and polarization management of the various beams. L: laser, PLL: phase-locked loop, PD: photodetector, PDH: Pound-Drever-Hall, COB: central optical bench, EOM: electro-optic modulator.}
    \label{fig:rc_layout}
\end{figure}

The RC, shown in Figure \ref{fig:rc_layout}, is formed by two mirrors positioned 123 meters apart. The cavity mirrors are in separate cleanrooms located roughly 26\,m underground within the tunnel formerly occupied by the HERA accelerator \cite{hera}. The RC adopts a plano-concave cavity geometry, where the cavity eigenmode forms a waist at mirror $\rm M2$ and the other mirror $\rm M1$ is positioned approximately a Rayleigh length away. This geometry minimizes the size of the projected RC eigenmode within the combined production and regeneration magnet strings to reduce clipping losses of the final ALPS\;II design due to the limited aperture of the magnets. The mirror $\rm M2$ of the cavity is nominally flat, while the nominal radius of curvature of the curved mirror $\rm M1$ is 214\,m, yielding a design cavity g-parameter\footnote{$g_{1,2} = 1-L_{\!_{\rm RC}}/R_{1,2}$ with $L_{\!_{\rm RC}}$ referring to the RC length, and $R_{1,2}$ the radius of curvature of either $\rm M1$ or $\rm M2$.} of $g_1g_2 = 0.43$. This geometry results in a cavity eigenmode with a waist located at the flat mirror with radius $w_0 = 6.0$\,mm and a beam radius on the curved mirror of 9.2\,mm.

The mirrors consist of fused silica substrates with a specified surface roughness of $< 0.1\,\rm nm$ (RMS), measured over $\rm 1\,mm \times 1\,mm$ sections of the mirror. The substrates are coated via ion-beam sputtering with alternating dielectric layers of tantala (Ta$_2$O$_5$) and silica (SiO$_2$). The flat and curved cavity mirrors are coated to have nominally specified transmissivities of 6.7 ppm and 107 ppm, respectively, at our operational wavelength of 1064 nm. The curved cavity mirror is mounted atop a precision actuation stage to allow for sub-$\murm$rad angular adjustments of the optic. The flat cavity mirror is rigidly mounted to the central optical bench (COB), an aluminum breadboard. The COB is placed on an alignment system which provides sub-$\murm$m positional and sub-$\murm$rad angular adjustment capability in six degrees of freedom.  

An ultra-high vacuum ($< 10^{-8}$ mbar) system encloses the entire RC, including vacuum chambers containing each of the mirrors, connected by the magnet bore. The twelve superconducting dipole magnets of the regeneration region, originally curved in order to fit the circular HERA accelerator ring, were mechanically straightened in order to provide a sufficient aperture for the fundamental cavity eigenmode. This process yielded dipoles with horizontal apertures between 46 and 51 mm \cite{albrecht2021magnets}. In order to minimize the risk of intra-cavity losses from clipping of the eigenmode, the magnets were aligned to a theoretical optical axis centered on the two cavity mirrors to within 200 $\murm$m. Where possible, the individual magnets are ordered such that those with the largest measured apertures are located where the beam is largest. The diameter of the cavity mirrors was chosen to be 50.8\,mm for the flat mirror and  75.2\,mm for the curved mirror such that the mirrors themselves would not limit the free aperture of the system.

The detector end station is equipped with two Nd:YAG nonplanar ring oscillator (NPRO) lasers with output of up to 500\,mW of continuous 1064\,nm light, while the central station has a single 200\,mW NPRO also with a wavelength of 1064\,nm. Most of the power from each laser is diverted, such that less than 10\,mW of power is incident on the RC. The detector end station lasers (L1 and L2) are spatially combined via a polarization-maintaining fiber beamsplitter and mode-matched and injected to the cavity using a series of lenses and mirrors. The frequency of L1 can be stabilized to the RC resonance using a Pound-Drever-Hall (PDH) control technique \cite{drever1983laser, black2001introduction} with a fiber electro-optic modulator (EOM) on the injection path and a photodetector in reflection of the cavity, PD1. The frequency of L2 can be controlled relative to L1 with an offset phase-locked loop (PLL) \cite{schunemann1999} using the beat-note formed on a photodetector (PD2) at one output of the fiber beamsplitter.
The central station laser, L3, is similarly mode-matched and injected into the cavity via the flat mirror $\rm M2$. The frequency of L3 can also be controlled relative to L1 in transmission of the RC using a PLL using the beat-note from the AC-coupled port of the cavity transmission photodetector, PD3. A combination of analog and digital-based servo instruments are used to control the laser frequencies, with stable lock periods on the scale of days. PD1, PD2, and PD3 are all identical high-speed (150\,MHz bandwidth, 2.3\,ns rise time) amplified photodetectors. 

\section{Measurement Techniques}
As Equations \ref{eq:signalrate} and \ref{eq:powerbuildup} show, the sensitivity of the ALPS\;II experiment relies on having accurate knowledge of the RC individual mirror transmissivities, as well as the round-trip losses. These quantities are obtained using two distinct methods - first, using the conventional cavity ring-down measurement and then by measuring the complex reflectivity coefficient of the cavity.

\subsection{Cavity ring-down}

\begin{figure}
\centering
    \begin{subfigure}[t]{.49\textwidth}
        \centering 
        \includegraphics[width=\textwidth]{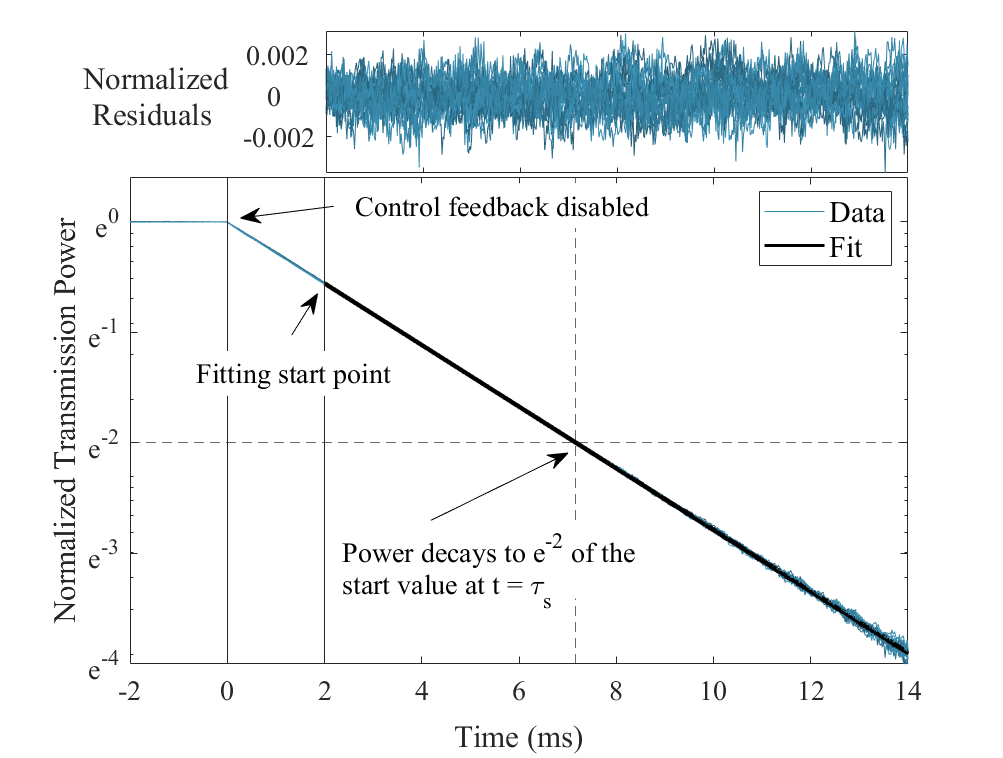}
        \caption{}
    \end{subfigure}
    \begin{subfigure}[t]{.49\textwidth}
        \centering
        \includegraphics[width=\textwidth]{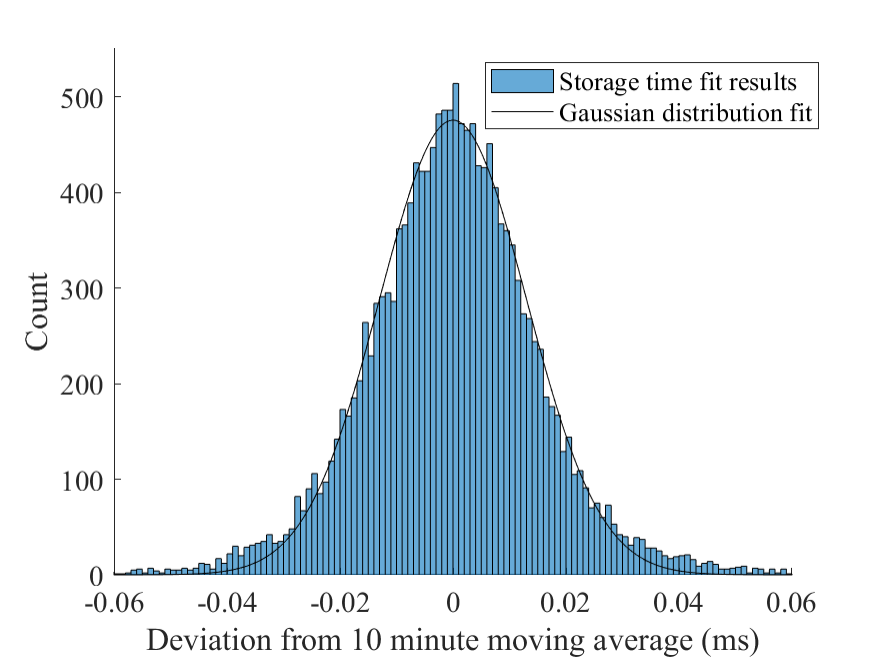}
        \caption{}
\end{subfigure}
\caption{a.) Measurement of the exponential decay of the power in transmission of the RC after disabling the feedback control, normalized to the power in the steady locked state. Twenty raw measurements are shown overlaid in shades of blue, along with unique corresponding lines of best fit in black. Each measurement was fit using nonlinear regression to the model $a + b e^{-2t / \tau_{\rm s}}$, yielding a storage time of $\tau_s = 7.17 \pm 0.01  \; \mathrm{ms}$. Above, the fit residuals are shown. b.) Histogram of 13,000 storage time deviations from a 10 minute moving average taken over 72 hours. A Gaussian distribution is fit to the data, with a standard deviation of 0.012\,ms.}
\label{fig:ringdown}
\end{figure}

We first characterized the RC by measuring the power of light in reflection and transmission of the cavity as the input field goes from a state of being stably on-resonance with the RC fundamental mode to far off-resonance. We partition the total input $P_{\rm in}$ into the powers in the fundamental spatial cavity mode ($P_0$) and in higher-order modes ($P_1$), such that the spatial mode-matching is $\eta_{\!_{\rm \, RC}} = P_0 / (P_0 + P_1)$. The power in transmission of the cavity before and after the input field is made to be non-resonant at time $t = 0$ is
\begin{align} \label{eq:ringdown}
    P_\mathrm{trans}(t) = 
    \begin{cases}
        P_0 \beta T_2, & t\leq 0.\\
        P_0 \beta T_2 e^{-2t / \tau_\text{s}}, & t \geq 0.
  \end{cases}
\end{align}
and in reflection,
\begin{align}\label{eq:ringdownreflection}
    P_\mathrm{ref}(t) \approx 
    \begin{cases}
        P_0 \left(1 - \beta(T_{\rm 2} + l) \right) + P_1 , & t<0.\\
        P_0 \left(1 + \beta T_1 e^{-2t / \tau_\text{s}}\right) + P_1, & t \geq 0.
  \end{cases}
\end{align}
From these equations, we are able to obtain values for the input mirror transmissivity $T_1$, the storage time $\tau_{\rm s}$, and the RC coupling efficiency $\eta_{\!_{\rm \, RC}}$ from which other cavity parameters can be derived \cite{isogai13}. 

We began with L1 frequency-stabilized to be on resonance with the RC fundamental mode using the curved mirror $\rm M1$ as the input mirror. The cavity transmitted power was measured on photodetector PD3 and recorded on a digital oscilloscope. Additionally, the power of light reflected from the cavity was simultaneously measured on an identical photodetector. To measure cavity ring-down in reflection and transmission, frequency control to the laser was turned off, such that L1 was no longer resonant with the RC. This process took less than $10\,\murm \rm s$, verified by measuring the duration of the sharp step-like response of the laser power in reflection described in Equation \ref{eq:ringdownreflection}.

The cavity reflected and transmitted powers were measured as they exponentially decay in accordance with Equation \ref{eq:ringdown}, and the collected time-series were fit with an exponential of the form $a + b e^{-2t / \tau_{\rm s}}$. The exponential was fit using a MATLAB nonlinear least-squares regression model. In each fit, the first two milliseconds after control was disengaged were excluded. Occurrences in which the laser frequency drifts back over the same, or another, cavity resonance within the measurement time were distinguishable in the transmitted power, and were systematically excluded using a root-mean-square error threshold on the fit. Approximately 80\% of measurements satisfied these criteria. Figure \ref{fig:ringdown}a shows a plot of twenty such accepted ring-down measurements. Figure \ref{fig:ringdown}b shows the statistical uncertainty of the ring-down measurement and fitting procedure from a set of 13,000 fits, demonstrating high consistency between sequential measurements. A value for the input mirror transmissivity $T_1$ was also obtained using laser power in reflection of the RC following from Isogai \textit{et al.} \cite{isogai13}, and reported in Table \ref{tab:results}.

\subsection{Cavity complex reflectivity}

We employed a technique of measuring the cavity's complex reflectivity as a function of frequency using heterodyne interferometry \cite{spector2024}.
The cavity complex reflectivity $\mathcal{R}$ is given by the ratio of the reflected and incident fields. For a high-finesse cavity, 
this can be approximated as
\begin{equation}
     \mathcal{R}(\Delta \nu) \equiv \frac{E_{\rm ref}}{E_{\rm in} } \approx  1- \displaystyle\frac{T_{\rm in}}{\displaystyle\frac{1}{2}(T_{\rm in} + T_{\rm out} + l) - 2\pi i\displaystyle\frac{\Delta\nu}{f_0}},
    \label{Eq:E_r}
\end{equation}
where $\Delta\nu$ is the frequency difference between the input laser field and the nearest cavity resonance, $T_{\rm in}$ is the transmissivity of the mirror from which the input field is reflected, and $T_{\rm out}$ is the transmissivity of the other mirror. 
Our experimental setup allows us to measure the complex reflectivity of the RC using the interference beat-note formed between a local oscillator, which transmits through the cavity, and a probe field, which can be scanned across resonance. Measuring the amplitude and phase of this beat-note yields the complex cavity reflectivity.

\begin{figure}
\centering
  \begin{subfigure}[t]{.49\textwidth}
  \centering
  \includegraphics[width=\textwidth]{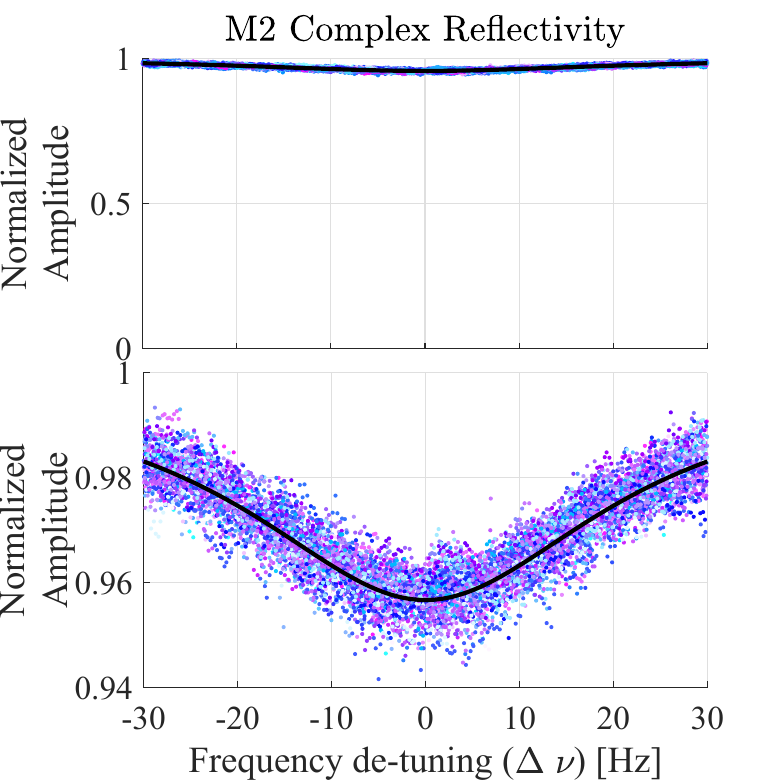}
  \caption{}
  \end{subfigure}
\hfill
\begin{subfigure}[t]{.49\textwidth}
    \centering
    \includegraphics[width=\textwidth]{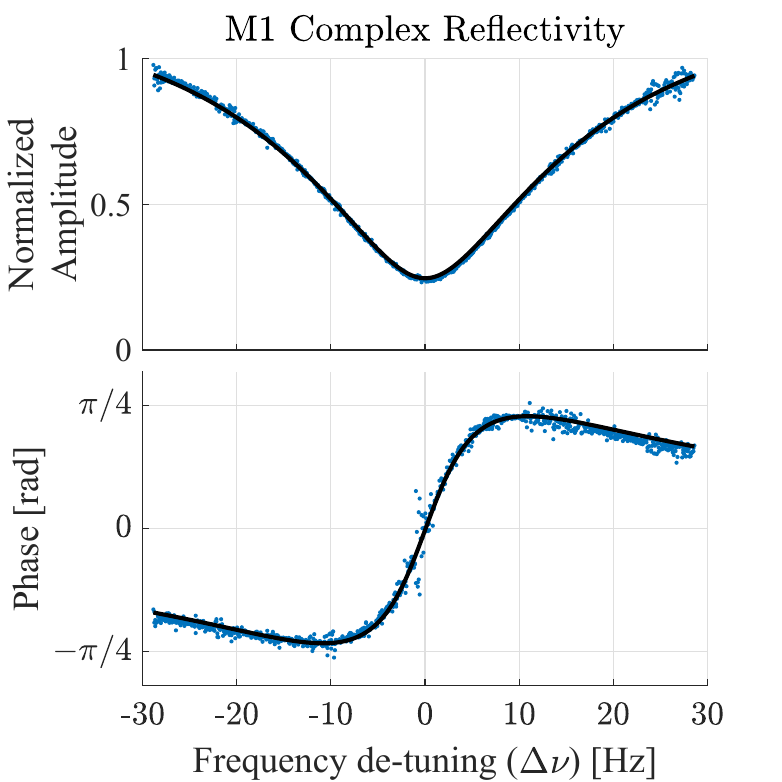}
    \caption{}
\end{subfigure}
    \caption{(a) Amplitude of the L1-L3 beat-note in reflection of the flat cavity mirror $\rm M2$ as L3 is scanned across resonance. Because of the low transmissivity of $\rm M2$ which results in poor signal-to-noise in the measurement, fifty scans (in shades of blue) are averaged. The amplitude component of Equation \ref{Eq:E_r} is fit (black) to the averaged data, which follow the cavity's complex reflectivity, yielding a transmissivity $T_2 = 5.1 \pm 0.5 \, \mathrm{ppm}$. The measurement is shown on the same scale as the $\rm M1$ amplitude measurement (top) and zoomed in (bottom). (b) Amplitude and phase of a single measurement of the L2-L3 beat-note in reflection of the curved cavity mirror $\rm M1$ as the probe field L3 is scanned across resonance. The complex reflectivity is again fit (black) to the amplitude and phase of the measured data, giving a transmissivity $T_1 = 95.7 \pm 0.5 \, \mathrm{ppm}$. The combination of both mirror reflectivity measurements and the cavity storage time is used to calculate excess optical losses of $142\pm1\, \mathrm{ppm}$ for the cavity alignment used in the ALPS\,II first science campaign. 
    }
    \label{fig:amp_phase}
\end{figure}

The RC reflectivity from the side of the flat mirror $\rm M2$ was interrogated by stabilizing the frequency of L1 to the  RC resonance, and offset phase-locking L3 relative to L1 by an integer number of free spectral ranges, such that the frequency of L3 could be scanned via the PLL offset across resonance. A frequency response analyzer was used to scan the PLL offset frequency over a cavity linewidth, centered around an integer number of free spectral ranges, and measured the amplitude of the L1-L3 beat-note formed at that offset frequency on PD3, shown in Figure \ref{fig:amp_phase}a. The phase information of the RC complex reflectivity in reflection of $\rm M2$ is obscured by the L1-L3 PLL, which works to actively suppress phase variations between the two fields, but the amplitude response can still be used to determine a value for $T_2$.

A measurement of the complex cavity reflectivity as seen from the side of the curved mirror $\rm M1$ was performed by bringing L1 onto RC resonance and phase-locking L3 to the transmitted L1 field, offset by an integer number of free spectral ranges. This resulted in L1 and L3 becoming simultaneously resonant within the RC, each injected via opposite cavity mirrors. L2, which is offset phase-locked to L1, was then scanned in frequency over a cavity resonance. The amplitude and phase of the beat-note formed by the interference between the cavity-transmitted L3 and the scanning L2 on PD1 was used to obtain the RC complex reflectivity using the curved mirror as the input mirror. Figure \ref{fig:amp_phase}b shows the amplitude and phase of the L2-L3 beat-note as a function of frequency around resonance, yielding the cavity linewidth as well as the transmissivity of the curved cavity mirror.

\section{Results}
In Table \ref{tab:results} we report the cavity parameters measured by the two techniques described earlier in this paper. The RC storage time at its maximum value was found to be $7.17 \pm 0.01 \, \rm ms$, to our knowledge more than 30\% longer than any previously reported value \cite{DellaValle2014}. Utilizing the cavity's complex reflectivity, we have measured the transmissivity of each mirror, in-situ and with the relevant cavity eigenmode position, to 0.5\,ppm precision. Both procedures are noninvasive to the ALPS\,II experiment, allowing us to take frequent measurements of the cavity performance over time and track changes during the first science campaign. 

\begin{table}
    \centering
    \begin{tabular}{l c c c} 
     \hline \hline
     \makecell{Cavity parameter} & \makecell{Complex Reflectivity \\(Science Runs)} & \makecell{Ring-down \\(Science Runs)} & \makecell{Ring-down \\(Lowest loss)} \\ 
     \hline
     Length ($L_{\rm RC}$) & $122.6012\pm0.0001\,\rm m$ & -- & -- \\ 
     Finesse ($\mathcal{F}$) & $25850\pm 100$ & $25650\pm 50$ & $27550 \pm 50$ \\
     Storage Time ($\tau_{\rm s}$) & $6.73\pm0.03\,\rm ms$ & $6.68 \pm 0.01 \rm\,ms$ & $7.17 \pm 0.01 \rm\,ms$ \\
     M1 Transmissivity ($T_1$) & $95.7\pm0.5\,\rm ppm$ & $96^{+1}_{-4}\,\rm ppm$ & $99^{+1}_{-4} \rm\, ppm$ \\
     M2 Transmissivity ($T_2$) & $5.1\pm 0.5\,\rm ppm$ & -- & -- \\
     Round-trip Losses ($l$) & $142\pm1\,\rm ppm$ & $144^{+4}_{-1}\rm\,ppm$ & $124^{+4}_{-1}\rm\,ppm$ \\
     Resonant Enhancement ($\beta$) & $6480\pm 50$ & $6400^{+60}_{-270}$ & $7610^{+75}_{-300}$ \\
     \hline
    \end{tabular}
    \caption{Summary of RC parameters derived from complex reflectivity measurements at the operational alignment for the ALPS\,II first science campaign, as well as cavity ring-downs performed at the operational alignment and the alignment associated with our minimum observed optical loss. For the ring-down parameter derivations, the values for length and $T_2$ from the complex reflectivity measurement are assumed.}
    \label{tab:results}
\end{table}

For the results presented in the first two columns of Table \ref{tab:results}, the RC was aligned such that a round-trip optical loss of 144\,ppm was measured for the fundamental eigenmode. As will be discussed in Section \ref{mirrormaps}, this configuration adequately represents the state of the RC during the first ALPS II science campaign and the values measured with the complex reflectivity method agree with simultaneously performed ring-down measurements. The cavity ring-down measurements reported in the third column represent the highest cavity storage time we were able to achieve by scanning the cavity fundamental eigenmode position on the mirrors.

\subsection{Mirror loss spatial distribution}\label{mirrormaps}

\begin{figure}
    \centering
    \includegraphics[width=\textwidth]{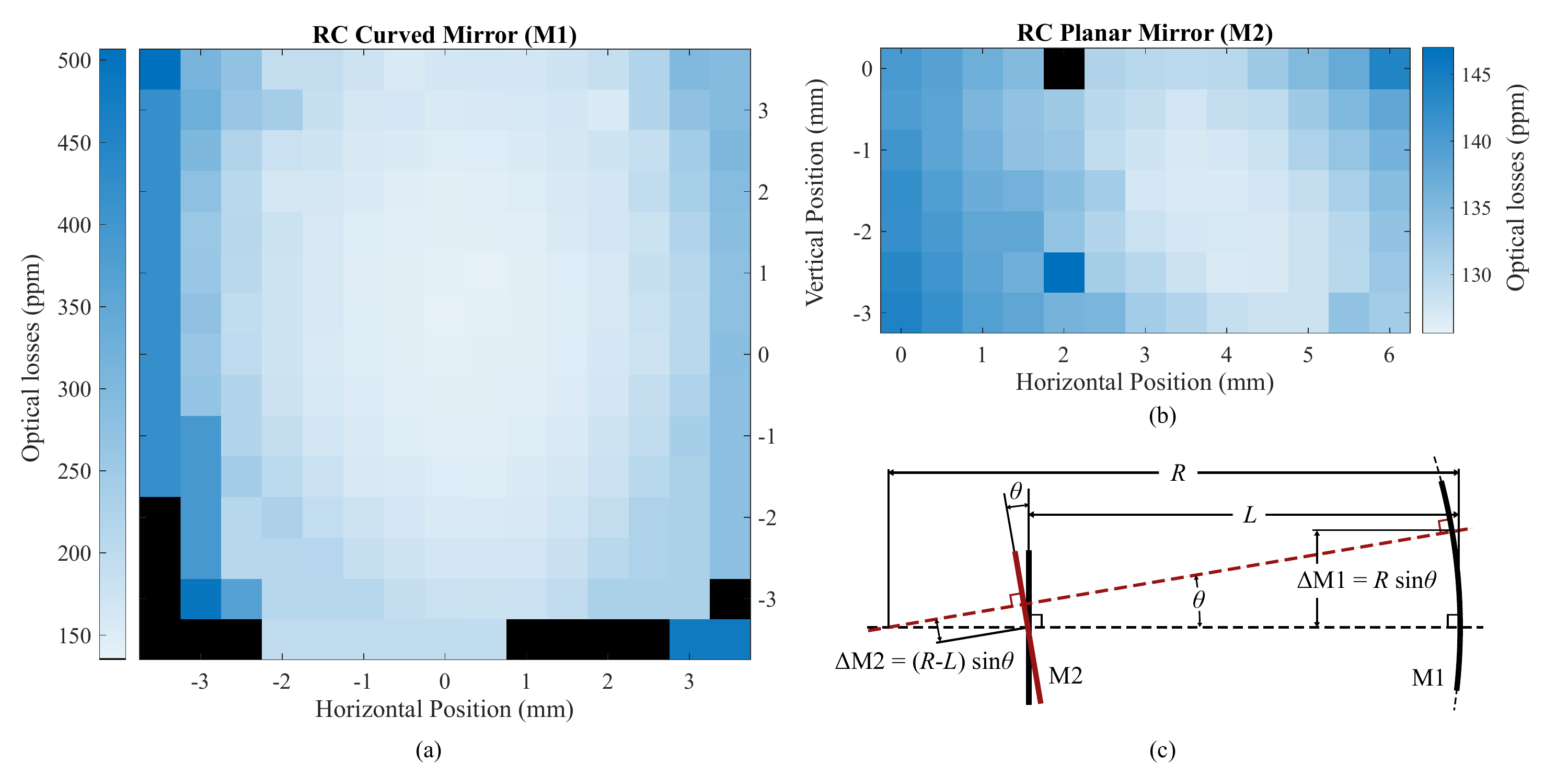}
    \caption{RC round-trip optical losses as a function of the relative position of the fundamental eigenmode for the RC curved mirror $\rm M1$ (a) and planar mirror $\rm M2$ (b). Black squares represent invalid measurements. The geometry of the cavity depicted in (c) shows how the cavity axis moves on each mirror as $\rm M2$ is tilted by an angle $\theta$ as a function of the cavity length $L$ and M1 radius of curvature $R$.}
    \label{fig:ringdown_map}
\end{figure}

The optical losses as a function of the cavity eigenmode position on the mirrors was also investigated. The distributions of optical losses are typically found to be non-uniform across even the highest-quality cavity mirror surfaces \cite{isogai13,gutierrez2023,Cui2017,Straniero15}. We measured the loss distribution of our mirrors, seen in Figure \ref{fig:ringdown_map}, by altering the location of the circulating fundamental cavity eigenmode on the optics and measuring the round-trip losses at each position. The maps represent the optical losses integrated over the Gaussian beam profile. Therefore, the spatial resolution of the maps are limited by the size of the eigenmode on the flat and curved cavity mirrors. The RC fundamental eigenmode is found to be slightly astigmatic, with a waist measuring 7.2 mm in the horizontal axis and 6.6 mm in the vertical axis on the flat cavity mirror (see Section \ref{Sec:Geo}).

To scan the eigenmode over the flat cavity mirror $\rm M2$, the position of the central optical bench was laterally translated with respect to the cavity optical axis. A 3\,mm $\times$ 6\,mm rectangular section of the mirror was investigated in steps of 0.5\,mm, limited by the safe actuation range of the alignment system for the COB. A CCD camera in transmission of the cavity was used to track the beam position and ensure that the actuators cause no unintended angular tilts of the flat cavity mirror. At each position of the COB, twenty ring-down measurements were made and the mean of the fit storage times which passed our selection criteria is reported in Figure \ref{fig:ringdown_map}b.

A loss map of the RC curved mirror $\rm M1$, shown in Figure \ref{fig:ringdown_map}a, was performed by executing angular tilts of the COB, causing the cavity eigenmode to translate on both $\rm M1$ and $\rm M2$ as well as through the beam tube in accordance with the illustration in Figure \ref{fig:ringdown_map}c. For a tilt of $\rm M2$ by angle $\theta$, the lateral translations of the beam spot position on each mirror can be derived geometrically as:
\begin{align}
\Delta \mathrm{M1}  = R \cdot \sin(\theta) && \Delta \mathrm{M2}  = (R - L) \cdot \sin(\theta)
\end{align}
where $R$ is the radius of curvature of $\rm M1$ and $L$ is the RC length.

Since the variation in losses observed from the scan of $\rm M1$ are far in excess of those from $\rm M2$, this map can be interpreted as a map of the combined losses from $\rm M1$ as well as clipping losses from the beam tube free aperture near the detector end station, with an uncertainty of 10\,ppm due to the fact that the shift of the eigenmode location on $\rm M2$ was not corrected for. The COB was tilted by actuating on a linear motor in steps of 2.5\,$\murm$m on one end of a 1.075\,m lever arm, yielding an angular resolution of 2.3\,$\murm$rad. Using an approximate calibration of 0.2\,mm/$\murm$rad for the eigenmode translation on $\rm M1$, an area of $\sim$ 7\,mm $\times$ 7\,mm of the curved cavity mirror was scanned.

A maximum storage time of 7.17\,ms was measured, corresponding to a minimum round-trip optical loss of 124 ppm. The twenty ring-down measurements shown in Figure \ref{fig:ringdown} are performed at this point. Areas of much higher optical loss are found near the edges of the curved cavity mirror scan, which we can associate with losses due to clipping on the aperture of the dipole magnet string. Modelling the clipping as resulting from  a pair of off-center circular apertures, we estimate a free aperture at the position of the magnet closest to the curved mirror of $42\pm 1$\,mm in the horizontal and $55 \pm 1$\,mm in the vertical direction. The vertical free aperture is consistent with the 55\,mm inner diameter of the individual magnet beam pipes, while the horizontal aperture is smaller than individual magnets, suggesting a possible misalignment of the magnets in the horizontal direction. When the fundamental cavity eigenmode is centered to the aperture, the cavity optical losses due to clipping are believed to be between 15 and 20\,ppm, with the uncertainty related to not knowing the exact eccentricity of the cavity eigenmode at the location of the clipping (see Section~\ref{Sec:Geo}).

During the ALPS\,II first science campaign, we selected an operational alignment which was regularly maintained. This alignment correspond to the origin (0,0) in both the flat and curved cavity mirror maps. The selection of eigenmode position on the flat mirror $\rm M1$ was constrained in the ALPS\,II first science campaign by alignment requirements on the central optical bench, as well as the height actuation limit of the COB alignment system. In the lateral COB position associated with the lowest measured optical losses, L3 could not be aligned to the RC. The cavity at this operational position was characterized using both ring-down measurements and the complex reflectivity, and the results are reported in Table \ref{tab:results}.

\subsection{Loss and coupling over time}\label{longterm}

\begin{figure}
    \centering
    \includegraphics[width=\textwidth]{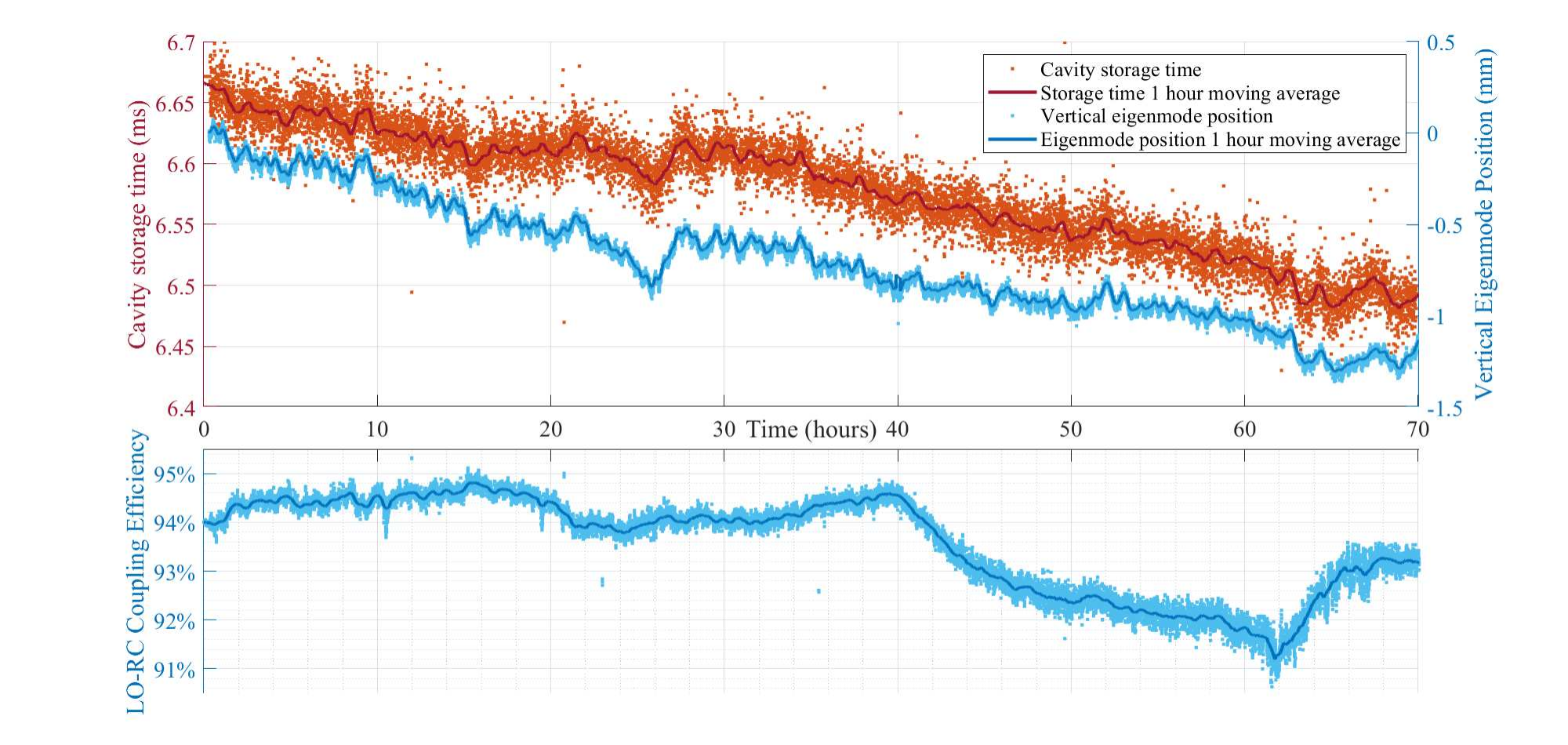}
    \caption{RC storage time $\tau_{\rm s}$ (top, red) as a function of time, along with a one-hour moving average. The right-hand y-axis shows the relative vertical eigenmode position at the location of the flat cavity mirror (top, blue) and a one-hour moving average. 
    On bottom, the cavity coupling efficiency $\eta_{\!_{\rm \, RC}}$ is shown for the same period of time.}
    \label{fig:storagetime_timeseries}
\end{figure}

Figure \ref{fig:storagetime_timeseries} shows the RC storage time $\tau_{\rm s}$ (top, red) measured by cavity ring-down approximately every 20 seconds over a three day period, along with a one-hour moving average. The vertical position of the eigenmode position with respect to a CCD camera fixed to the optical table in transmission of the cavity (top, blue) shows very strong correlation to the storage time. Practically no changes in the horizontal position were observed. This suggests that much of the variation in the storage time can be explained by motion of the fundamental eigenmode position on the mirrors due to natural alignment changes, primarily in the vertical direction.

The bottom part of Figure \ref{fig:storagetime_timeseries} shows the coupling efficiency between the RC and L1, $\eta_{\!_{\rm \, RC}}$, over the same time period. This cavity coupling was calculated using the power in reflection of the cavity during locked and unlocked states, along with the storage time measured at each point. The coupling efficiency variations are loosely correlated with several environmental factors, in particular the temperature inside the laser cleanrooms. During the ALPS\,II science campaign, the coupling efficiency was monitored in this way and realignment of the input optics was manually performed daily if the coupling was found to be below a threshold of 90\%.

\subsection{Cavity length and geometry measurements}
\label{Sec:Geo}

The reflectivity of the RC was also used to determine the cavity free spectral range and higher-order mode spacing. To find the RC free spectral range, the frequency of L1 was stabilized to the cavity fundamental resonance and the L1-L2 PLL offset frequency was scanned until a sharp decline in the amplitude of the L1-L2 beat-note in reflection of the cavity was observed. The offset frequency where the beat-note amplitude reached a minimum in the distribution described by Equation \ref{Eq:E_r}, corresponding to L1 and L2 becoming simultaneously resonant in the RC, is an integer number of free spectral ranges. With both lasers resonant with the RC fundamental mode, the frequency difference between the lasers gave a free spectral range of 1.222632\,MHz $\pm$ 1\,Hz. The free spectral range is $f_0 = c / (2L_{\rm RC})$,
which gives a cavity length of 122.6012\,m $\pm$ 0.0001\,m. Long-term variations of the cavity length are discussed in the following section.

By stabilizing the frequency of L1 to be on resonance with the first order Hermite-Gauss eigenmode ($\rm HG_{01 / 10}$) of the RC and tuning the L1-L2 offset frequency until L2 became resonant with a fundamental mode of the RC ($\rm HG_{00}$), we were able to measure the higher-order transverse mode spacing. This measurement yielded a higher-order mode spacing of between 263.3 and 264.3\,kHz for the vertically oriented $\rm HG_{10}$ mode, depending on the position of the eigenmode on the mirror. For the horizontal $\rm HG_{01}$ mode, the spacing was between 294.5 and 295.8\,kHz, again depending on the eigenmode position. 

From the RC mode frequency spacing measurements, we were able to calculate the cavity g-parameters directly, resulting in  $g_1g_2\simeq0.527\pm0.002$ for the vertical axis and $g_1g_2\simeq0.607\pm0.003$ for the horizontal axis. It should be noted that this is not only different from the nominal value of $g_1g_2=0.43$, but also the difference in values for the vertical and horizontal axes shows that the cavity is slightly astigmatic. Assuming simple astigmatism with the flat mirror showing no curvature, this would correspond to an eigenmode waist radius of 7.2\,mm in the horizontal axis and 6.6\,mm in the vertical axis, with the waist located at the flat cavity mirror. While the exact nature of the astigmatism is unknown, these projections roughly agree with measurements of the beam profile in transmission of the RC. Based on these measurements, the additional loss in mode-matching of an axially symmetric Gaussian beam to the cavity is on the order order of 1\% due to the astigmatism. This is corroborated by our ability to couple more than 95\% of the power of L1 to the cavity without the need for any components that help compensate for the astigmatism in the cavity eigenmode.

\subsection{Cavity length variations}
Changes in the RC length cause the cavity free spectral range $f_0 = c/(2L_{\rm RC})$ to also vary. This variation is measured using heterodyne interferometry between the lasers L1 and L2, both in transmission of the RC. As in the measurement of the static free spectral range described in the previous section, L1 is frequency stabilized to be on resonance with a RC fundamental eigenmode and the offset frequency of the L1-L2 PLL is tuned to be an integer number $n$ of free spectral ranges, such that L2 is also resonant with another fundamental eigenmode. Since the frequency of L1 is actively stabilized to resonance, but the frequency of L2 follows that of L1 by some fixed offset, any variation of the cavity length - and therefore the free spectral range - will cause a slight de-tuning of L2 from resonance. The change in frequency of the RC resonance relative to L2 is given by
\begin{equation}
    \Delta \nu = n \cdot f_0 \cdot \frac{\Delta L_{\rm RC}}{L_{\rm RC}}
\end{equation}
where $\Delta L$ is the change in the absolute cavity length.

The magnitude of the frequency de-tuning of L2 from resonance is measured in the phase of the heterodyne beat-note that they form on PD3. At this detector, in transmission of the cavity, both fields are subjected to the complex transmissivity of the RC. Similar to the complex reflectivity described in Equation \ref{Eq:E_r}, the cavity complex transmissivity $\mathcal{T}(\Delta \nu)$ is the ratio of the transmitted and incident fields. For high-finesse cavities,
\begin{equation}
    \mathcal{T}(\Delta \nu) \equiv \frac{E_{\rm trans}}{E_{\rm in} } \approx \frac{\sqrt{T_{\rm in}}\sqrt{T_{\rm out}}}{\displaystyle\frac{1}{2}(T_{\rm in} + T_{\rm out} + l) - 2\pi i\displaystyle\frac{\Delta\nu}{f_0}},
    \label{Eq:E_t}
\end{equation}
The phase of L2 is therefore shifted relative to L1 in transmission of the RC as the cavity length varies. For frequencies within the fundamental resonance linewidth, the phase of the complex transmissivity as a function of $\Delta L_{\rm RC}$ becomes nearly linear, and can be approximated in radians as
\begin{equation}
    \Delta \phi \approx - 2 \mathcal{F} \cdot n \cdot \frac{\Delta L_{\rm RC}}{L_{\rm RC}}
\end{equation}
where $\mathcal{F}$ is the RC finesse. This equation describes a slope which is tangent to the phase component of Equation \ref{Eq:E_t} at $\Delta \nu = 0$. The phase variations of the L1-L2 beat-note are recorded using a digital phasemeter.

\begin{figure}
    \centering
    \includegraphics[width=\linewidth]{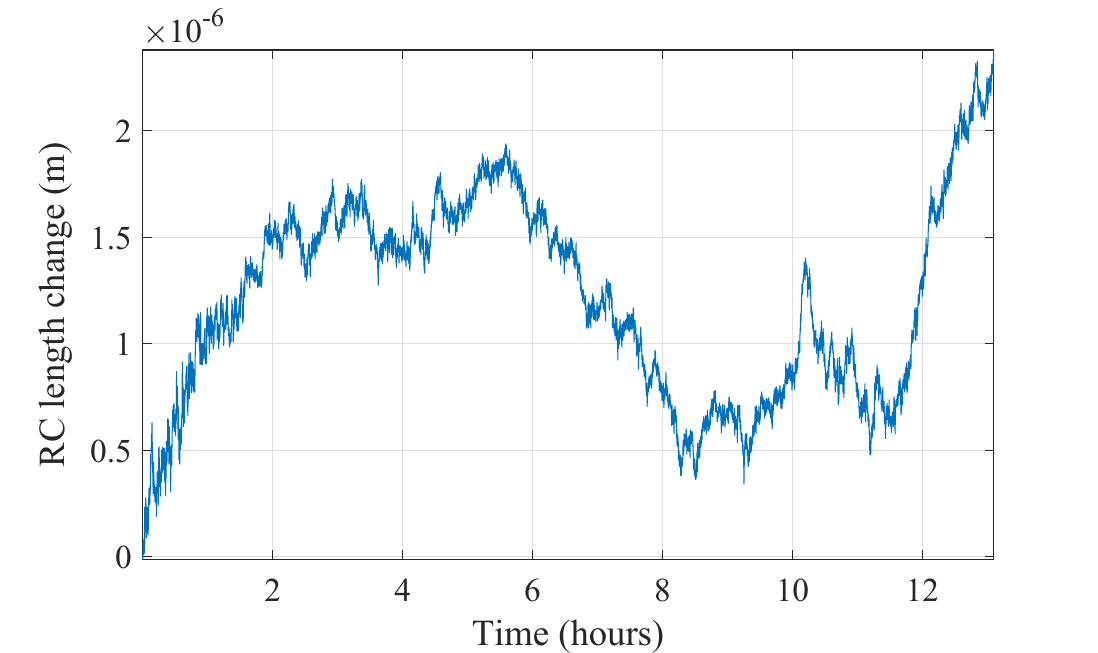}
    \caption{Time series of RC length variations measured over a continuous 13 hour period. The data is low-pass filtered at 0.1 Hz to remove unsuppressed noise in the L1-L2 PLL.}
    \label{fig:seismic_spectrum}
\end{figure}

Figure \ref{fig:seismic_spectrum} shows the phase variations of the L1-L2 beat-note in transmission of the RC, calibrated to cavity length in units of meters. The measurement was performed over a typical 13 hour overnight span, using a L1-L2 PLL frequency separation of $n = 45$ free spectral ranges. Above approximately 0.1\,Hz, the measurement noise is primarily from residual phase noise of the L1-L2 PLL, whereas below 0.1\,Hz the variations result from physical cavity length changes. The magnitude of the slow length variations is consistent with measurements of the environmental conditions of the experimental site made using accelerometer and seismometer data prior to site commissioning \cite{Miller2019}. During the ALPS\,II first science campaign, the RC free spectral range is routinely measured and compensated to minimize de-tuning of the high-power laser from RC resonance due to slow cavity length changes. In order to meet the requirement that the high-power laser not de-tune from RC resonance by more than 10\% of the linewidth, the RC length must not vary by more than 10\,$\murm$m between checks. The magnitude of the variations shown in Figure \ref{fig:seismic_spectrum}, less than 3\,$\murm$m peak-to-peak over the course of 13 hours, indicates that active length compensation is not necessary on these time scales. This measurement technique allows the RC to act as a strainmeter with a sensitivity on the order of $10^{-9} - 10^{-10}$ at low frequencies, comparable to similarly-sized dedicated Fabry-Perot strainmeters \cite{takamori2014}.

\begin{figure}
    \centering
    \includegraphics[width=\linewidth]{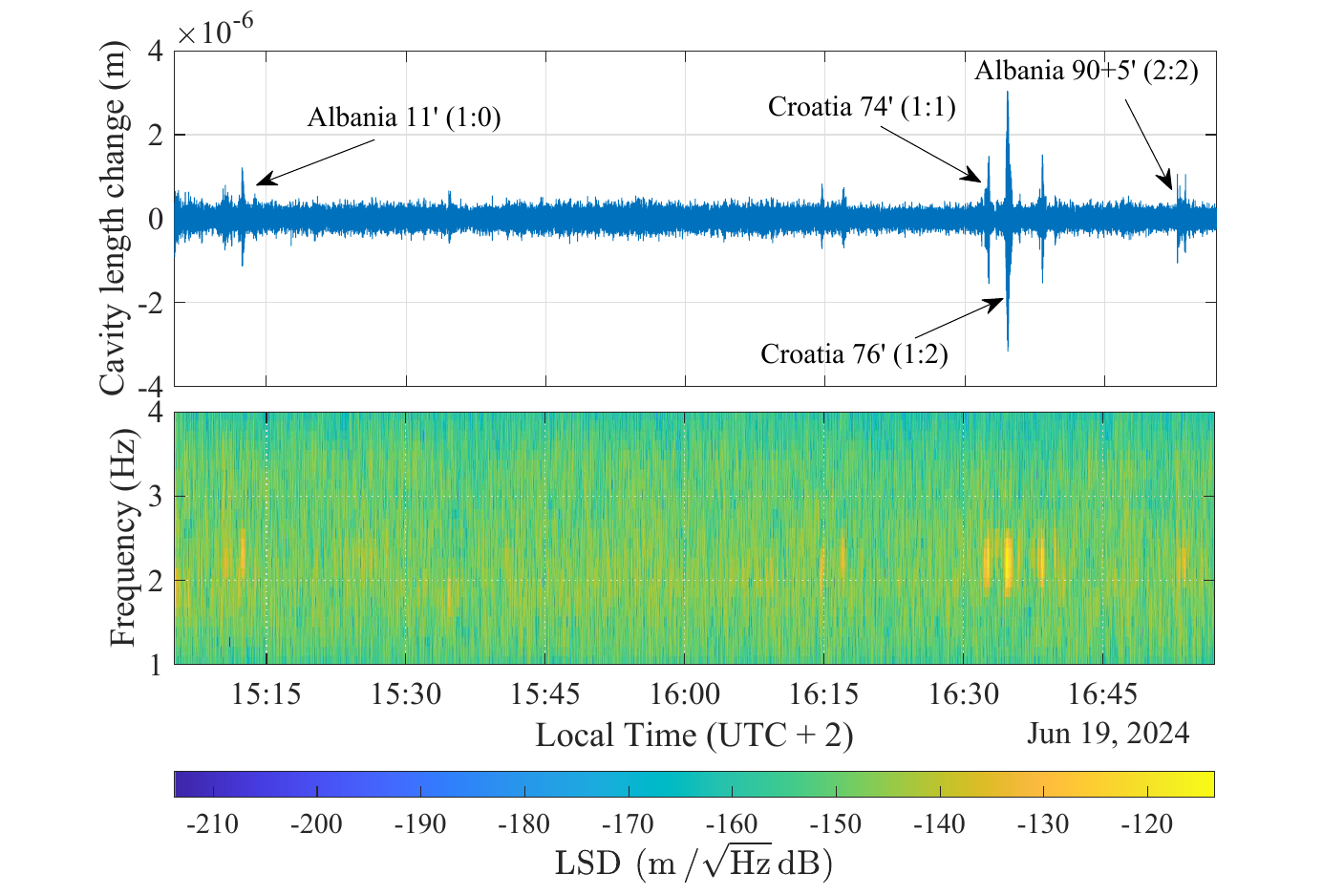}
    \caption{Time series band-pass filtered between 1 and 5 Hz (top) and amplitude spectrogram (bottom) of RC length variations over the course of the Albania-Croatia football match of the 2024 UEFA European Football Championship. Peaks associated with goals scored during the game are indicated by arrows labelled with the scoring team and game time.}
    \label{fig:football_noise}
\end{figure}

Using this absolute length sensing measurement, the RC is also sensitive to seismic signatures generated by human activity which generate cavity length changes larger than $\sim$ 100\,nm. In one example, shown in Figure \ref{fig:football_noise}, the cavity length was measured during a football match played in the Volksparkstadion, located 1\,km away from the experimental site, and attended by 47,000 people \cite{UEFA2024}. Immediately following each of the four goals scored during the game, crowd-induced RC length variations are observed, lasting 15-25 seconds with frequencies primarily in the 2 to 2.4\,Hz band in each case. The emergent character, frequency, and duration of these seismic transients are highly consistent with seismometer measurements of crowd-induced ``footquakes'' at football stadiums around the world \cite{Daz2017}\cite{Denton2018}. Ultimately, the short duration of these, or similar, anthropogenic seismic events do not affect the long-term length stability of the RC necessary for the ALPS\,II first science campaign, and the experiment can be operated even during major cultural events.

\section{Conclusion and Outlook}
We have characterized the ALPS\,II Regeneration Cavity, a 123 meter long Fabry-Perot optical cavity with a storage time as high as 7.17\,ms. In preparation for the ALPS\,II first science campaign, the round-trip optical losses of the cavity mirrors were considered as a function of position and time. This analysis yields a stable and predictable system which can be regularly measured during operation in order to calibrate the ALPS II experimental sensitivity. Using two complementary techniques, we have measured the input mirror transmissivity $T_1$ and the round-trip optical losses $l$, yielding a resonant enhancement factor $\beta$ of 6400 and a finesse of 25,000 in the alignment used for the first science campaign. This high resonant enhancement factor will contribute to making the ALPS\,II first science campaign the most sensitive light-shining-through-a-wall experiment ever performed.

The ALPS\,II RC resonant enhancement factor can be increased by improving the cavity optics. Of the optimized 124\,ppm of measured round-trip optical losses, approximately 20\,ppm can be attributed to clipping of the cavity fundamental eigenmode on the aperture of the magnet string. The remaining $\sim$100\,ppm of optical losses are higher than expected based on analysis of the mirror surface roughness. The remaining excess losses could be attributed to low spatial frequency surface error or particulate contamination of the optics. Each mirror had been handled and stored for several days in an optical cleanroom before entering vacuum. The installed mirrors also experienced cycles of venting the vacuum system with (clean) air, a process that has been associated with contamination and degraded mirror performance \cite{DellaValle2014}.

To further improve the ALPS\,II experimental sensitivity for future science campaigns, a second optical cavity will be installed in the production region. This production cavity will increase the power of light circulating in the magnetic field to increase the axion flux. The production cavity will have a matching plano-concave design to the RC, consisting of a planar mirror in the central station and a curved mirror in the laser end station, each coated identically to their RC counterparts. This configuration ensures that the two cavity fundamental eigenmodes have high spatial overlap $\eta$. The production cavity will be length-controlled in order to keep the cavity circulating field simultaneous on resonance with the RC, requiring a piezo-electric actuated mirror mount with a control bandwidth of 4 kHz \cite{pold2020}. Based on the seismic conditions at the experimental site observed by the RC, this design should be sufficient to suppress environmental noise in order to meet the requirements of the final ALPS\,II design \cite{diaz2022}.

\subsection*{Funding}
We gratefully acknowledge the support of the National Science Foundation (Grant no. PHY-2309918), the Heising-Simons Foundation (Grant no. 2020-1841), the Deutsche Forschungsgemeinschaft under Germany’s Excellence Strategy (EXC 2121 “Quantum Universe" 390833306 and WI 1643/2-1), the Partnership for Innovation, Education and Research (PIER) of DESY and Universität Hamburg under PIER Seed Project (PIF-2022-18), the German Volkswagen Stiftung, and the UK Science and Technologies Facilities Council (Grant no. ST/T006331/1).

\subsection*{Acknowledgments}
We are very thankful to all members of the ALPS\,II collaboration, as well as to the site and engineering support from the Deutsches Elektronen-Synchrotron DESY, in particular David Reuther, Sandy Croatto, Uwe Schneekloth, and the MKS and MVS divisions for their vacuum and cryogenic expertise. 

\subsection*{Disclosures} The authors declare no conflicts of interest.

\subsection*{Data availability} Data underlying the results presented in this paper are not publicly available at this time but may be obtained from the authors upon reasonable request.


\bibliography{rc}

\newcommand{\noop}[1]{}
\begin{thebibliography}{10}

\bibitem{Sikivie1983}
P.~Sikivie.
\newblock Experimental tests of the "invisible" axion.
\newblock {\em Phys. Rev. Lett.}, 51:1415--1417, Oct 1983.

\bibitem{Anselm1985}
A.~A. Anselm.
\newblock {Arion $\leftrightarrow$ Photon Oscillations in a Steady Magnetic Field. (In Russian)}.
\newblock {\em Yad. Fiz.}, 42:1480--1483, 1985.

\bibitem{vanbibber1987}
K.~Van~Bibber, N.~R. Dagdeviren, S.~E. Koonin, A.~K. Kerman, and H.~N. Nelson.
\newblock Proposed experiment to produce and detect light pseudoscalars.
\newblock {\em Phys. Rev. Lett.}, 59:759--762, Aug 1987.

\bibitem{diaz2022}
M.~{Diaz Ortiz}, J.~Gleason, H.~Grote, A.~Hallal, M.T. Hartman, H.~Hollis, K.-S. Isleif, A.~James, K.~Karan, T.~Kozlowski, A.~Lindner, G.~Messineo, G.~Mueller, J.H. Põld, R.C.G. Smith, A.D. Spector, D.B. Tanner, L.-W. Wei, and B.~Willke.
\newblock Design of the alps ii optical system.
\newblock {\em Physics of the Dark Universe}, 35:100968, 2022.

\bibitem{Hoogeveen1990}
F.~Hoogeveen and T.~Ziegenhagen.
\newblock {Production and detection of light bosons using optical resonators}.
\newblock {\em Nucl. Phys. B}, 358:3--26, 1991.

\bibitem{Sikivie2007}
P.~Sikivie, D.~B. Tanner, and Karl van Bibber.
\newblock Resonantly enhanced axion-photon regeneration.
\newblock {\em Phys. Rev. Lett.}, 98:172002, Apr 2007.

\bibitem{Mueller2009}
Guido Mueller, Pierre Sikivie, D.~B. Tanner, and Karl van Bibber.
\newblock Detailed design of a resonantly enhanced axion-photon regeneration experiment.
\newblock {\em Phys. Rev. D}, 80:072004, Oct 2009.

\bibitem{arias2010}
Paola Arias, Joerg Jaeckel, Javier Redondo, and Andreas Ringwald.
\newblock Optimizing light-shining-through-a-wall experiments for axion and other weakly interacting slim particle searches.
\newblock {\em Phys. Rev. D}, 82:115018, Dec 2010.

\bibitem{redondo2011}
Javier Redondo and Andreas Ringwald.
\newblock Light shining through walls.
\newblock {\em Contemporary Physics}, 52(3):211--236, 2011.

\bibitem{hera}
{\em {HERA - A Proposal for a Large Electron Proton Colliding Beam Facility at DESY}}.
\newblock 1981.

\bibitem{albrecht2021magnets}
Clemens Albrecht, Serena Barbanotti, Heiko Hintz, Kai Jensch, Ronald Klos, Wolfgang Maschmann, Olaf Sawlanski, Matthias Stolper, and Dieter Trines.
\newblock Straightening of superconducting hera dipoles for the any-light-particle-search experiment alps ii.
\newblock {\em EPJ Techniques and Instrumentation}, 8(5), 2021.

\bibitem{drever1983laser}
RWP Drever, John~L Hall, FV~Kowalski, J\_ Hough, GM~Ford, AJ~Munley, and H~Ward.
\newblock Laser phase and frequency stabilization using an optical resonator.
\newblock {\em Applied Physics B}, 31(2):97--105, 1983.

\bibitem{black2001introduction}
Eric~D Black.
\newblock An introduction to pound--drever--hall laser frequency stabilization.
\newblock {\em American journal of physics}, 69(1):79--87, 2001.

\bibitem{schunemann1999}
U.~Schünemann, H.~Engler, R.~Grimm, M.~Weidemüller, and M.~Zielonkowski.
\newblock {Simple scheme for tunable frequency offset locking of two lasers}.
\newblock {\em Review of Scientific Instruments}, 70(1):242--243, 01 1999.

\bibitem{isogai13}
T.~Isogai, J.~Miller, P.~Kwee, L.~Barsotti, and M.~Evans.
\newblock Loss in long-storage-time optical cavities.
\newblock {\em Opt. Express}, 21(24):30114--30125, Dec 2013.

\bibitem{spector2024}
Aaron~D. Spector and Todd Kozlowski.
\newblock Optical cavity characterization with a mode-matched heterodyne sensing scheme.
\newblock {\em Opt. Express}, 32(16):27112--27124, Jul 2024.

\bibitem{DellaValle2014}
F.~Della Valle, E.~Milotti, A.~Ejlli, U.~Gastaldi, G.~Messineo, L.~Piemontese, G.~Zavattini, R.~Pengo, and G.~Ruoso.
\newblock Extremely long decay time optical cavity.
\newblock {\em Opt. Express}, 22(10):11570--11577, May 2014.

\bibitem{gutierrez2023}
N.~Gutierrez, J.~Degallaix, D.~Hofman, C.~Michel, L.~Pinard, J.~Morville, R.~Battesti, and G.~Cagnoli.
\newblock {Optical characterization of high performance mirrors based on cavity ringdown time measurements with 6 degrees of freedom mirror positioning}.
\newblock {\em Review of Scientific Instruments}, 94(10):105113, 10 2023.

\bibitem{Cui2017}
Hao Cui, Bincheng Li, Shilei Xiao, Yanling Han, Jing Wang, Chunming Gao, and Yafei Wang.
\newblock Simultaneous mapping of reflectance, transmittance and optical loss of highly reflective and anti-reflective coatings with two-channel cavity ring-down technique.
\newblock {\em Opt. Express}, 25(5):5807--5820, Mar 2017.

\bibitem{Straniero15}
Nicolas Straniero, J\'{e}r\^{o}me Degallaix, Raffaele Flaminio, Laurent Pinard, and Gianpietro Cagnoli.
\newblock Realistic loss estimation due to the mirror surfaces in a 10 meters-long high finesse fabry-perot filter-cavity.
\newblock {\em Opt. Express}, 23(16):21455--21476, Aug 2015.

\bibitem{Miller2019}
Dominik Miller.
\newblock {\em Seismic noise analysis and isolation concepts for the ALPS II experiment at DESY}.
\newblock PhD thesis, Gottfried Wilhelm Leibniz Universität Hannover, 2019.

\bibitem{takamori2014}
Akiteru Takamori, Akito Araya, Wataru Morii, Souichi Telada, Takashi Uchiyama, and Masatake Ohashi.
\newblock A 100-m fabry–pérot cavity with automatic alignment controls for long-term observations of earth’s strain.
\newblock {\em Technologies}, 2(3):129--142, 2014.

\bibitem{UEFA2024}
Union of~European Football~Associations.
\newblock Full time report – croatia v albania.
\newblock \url{https://www.uefa.com/newsfiles/EURO/2024/2036176_FR.pdf}.

\bibitem{Daz2017}
Jordi D{\'i}az, Mario Ruiz, Pilar S{\'a}nchez‐Pastor, and Paula Romero.
\newblock Urban seismology: on the origin of earth vibrations within a city.
\newblock {\em Scientific Reports}, 7, 2017.

\bibitem{Denton2018}
Paul Denton, Stewart Fishwick, Victoria Lane, and Debra Daly.
\newblock {Football Quakes as a Tool for Student Engagement}.
\newblock {\em Seismological Research Letters}, 89(5):1902--1907, 06 2018.

\bibitem{pold2020}
{Põld, Jan H.} and {Spector, Aaron D.}
\newblock Demonstration of a length control system for alps ii with a high finesse 9.2 m cavity.
\newblock {\em EPJ Techn Instrum}, 7(1):1, 2020.

\end{thebibliography}

\end{document}